\documentclass[twocolumn,superscriptaddress,amsmath,amssymb,floatfix,aps,prl,showpacs]{revtex4}
\usepackage{graphics,graphicx}

\begin{document}

\title{Soft Spheres Make More Mesophases}
\author{Matthew A. Glaser}
\affiliation{Department of Physics, University of Colorado, Boulder, CO 80309-0390, USA}

\author{Gregory M. Grason}
\affiliation{Department of Physics and Astronomy, University of California at Los Angeles, Los Angeles, CA 90024, USA}

\author{Randall D. Kamien}
\affiliation{Department of Physics and Astronomy, University of Pennsylvania, Philadelphia, PA 19104, USA}

\author{A. Ko\v smrlj}
\altaffiliation{Present address: Massachusetts Institute of Technology, Cambridge, MA 02139-4307, USA.}
\affiliation{Department of Physics, University of Ljubljana, Jadranska 19, SI-1000 Ljubljana, Slovenia}

\author{Christian D. Santangelo}
\altaffiliation{Future address: Department of Physics, University of Massachusetts, Amherst, MA 01003-9337, USA.}
\affiliation{Department of Physics and Astronomy, University of Pennsylvania, Philadelphia, PA 19104, USA}

\author{P. Ziherl}
\affiliation{Department of Physics, University of Ljubljana, Jadranska 19, SI-1000 Ljubljana, Slovenia}
\affiliation{Jo\v zef Stefan Institute, Jamova 39, SI-1000 Ljubljana, Slovenia}

\date{\today}

\begin{abstract}

We use both mean-field methods and numerical simulation to study the phase diagram of classical particles interacting with a hard-core and repulsive, soft shoulder. Despite the purely repulsive interaction, this system displays a remarkable array of aggregate phases arising from the competition between the hard-core and shoulder length scales. In the limit of large shoulder width to core size, we argue that this phase diagram has a number of universal features, and classify the set of repulsive shoulders that lead to aggregation at high density.
Surprisingly, the phase sequence and aggregate size adjusts so as to keep almost constant inter-aggregate separation.

\end{abstract}
\maketitle

Entropy is a potent force in the theory of self-assembly. It can be argued, through entropic considerations alone, that hard spheres will self-assemble into the face-centered-cubic (fcc) lattice or any of its many variants related through stacking faults. As a result, when a material exhibits an fcc phase, it is often attributed to the optimality of the close-packed lattice. When less common or less dense lattices are formed, all sundry of explanations are invoked, ranging from quantum mechanics~\cite{Rabe}, lattice effects~\cite{AlexanderMcTague}, partially filled Landau levels~\cite{Shklovskii}, and even soft interactions~\cite{ZiherlKamienReview, GrasonReview}. While there has been concerted effort to tailor the pair interaction to achieve a desired periodic arrangement~\cite{Torquato} this must be done in the context of those packing motifs that arise from generic interactions. For instance, it would be no trick to tailor a potential to make an fcc lattice. 

\begin{figure}

\includegraphics{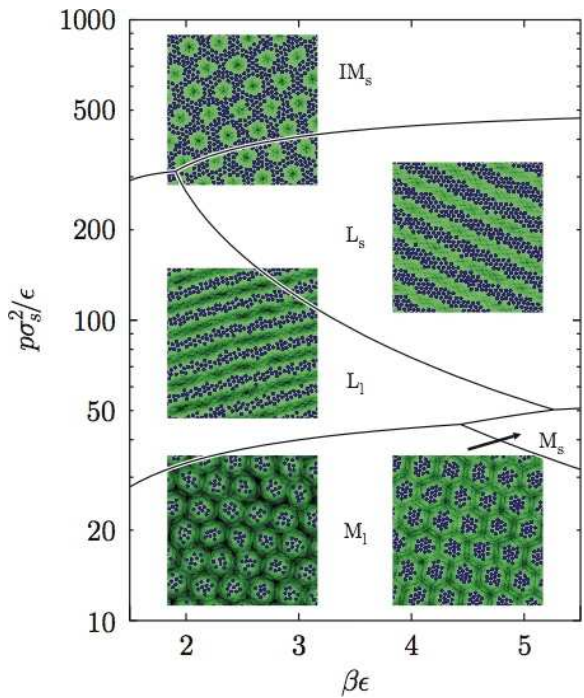}
\caption{Theoretical phase diagram and simulation snapshots of some solid and liquid modulated phases in two dimensions.  Shown are only the solid ($\rm M_s$) and liquid ($\rm M_l$) micelle phases, solid ($\rm L_s$) and liquid ($\rm L_l$) lamellar phases, and the solid inverse micelle phase ($\rm IM_s$).  Here, $\sigma_s/\sigma=5$ where $\sigma_s$ is the soft-shoulder radius and $\sigma$ the hard-core radius.  The potential height is $\epsilon$, the pressure is $p$, and $\beta = 1/k_BT$.  The dark circles are the particle hard cores and the diffuse circles represent the particle soft-shoulder.}
\label{fig:phasediagram}
\end{figure}

With this in mind, here we consider a seemingly simple extension of the hard sphere model, namely a hard core interaction of radius $\sigma$ with a soft shoulder of radius $\sigma_s$ and height $\epsilon$ (HCSS):
\begin{equation}
\label{eq:hcss}
V_{\rm HCSS}(r)=\left\{
\begin{array}{ll}
\infty& r<\sigma\\
\epsilon & \sigma<r<\sigma_s\\
0& r>\sigma_s
\end{array}
\right. 
\end{equation}
When $\sigma_s/\sigma\gtrsim 1$ this potential models hard spheres with a soft pair repulsion and was used to study isostructural transitions in Cs and Ce~\cite{Young77}. In generic repulsive potentials, it has been shown that as the range of the soft repulsion grows (corresponding to $\sigma_s/\sigma\approx 2$) a rich variety of density-modulated ground states appear~\cite{Malescio03,Camp03,Norizoe05} which can be characterized as periodic arrangements of regular sized clusters of the original spheres. In this Letter we establish a sufficient condition on the pair potential for clustering and the subsequent ordering of the clusters which generalizes results on soft potentials without hard cores~\cite{Likos01}. We develop a self-consistent field theory for soft repulsion and use it to study the formation of striped phases. We corroborate our analytic treatment with numerical solutions that also predict the existence of hexagonal and inverted hexagonal phases with both fluid and crystalline order in the clusters, as shown in Fig.~\ref{fig:phasediagram}.
We also present results from Monte Carlo simulations of the HCSS potential which both stimulate and support the more general results. In all cases, we find that over the range of stable aggregate structures the lattice constant remains fixed while the clusters change their size and morphology so as to maintain the average intra-cluster density $\rho$. 

Why should repulsive potentials lead to aggregation? We can understand this in the context of the HCSS model: consider a uniform density of spheres at densities just large enough so that the soft-shoulders only begin to touch the hard cores. The energy can be lowered by, for instance, bringing pairs of spheres together -- moving two spheres closer requires no additional energy, but by reducing the number of nearest neighbors the overall energy is reduced. To wit, the same physics drives the formation of a multiply-occupied crystal in the penetrable sphere model ($\sigma=0$)~\cite{Klein94}, while the ground state of a generalized exponential repulsion~\cite{Mladek06} is a multiply-occupied fcc lattice with a lattice constant independent of the average density. To explore the HCSS ensemble we introduce a lattice model with occupation $n_i=0,1$ at each site to enforce the hard core repulsion. The remaining soft shoulder is characterized by $V_{\rm SS}({\bf r})$ and the density is set through the chemical potential $\mu$ in the Hamiltonian:
\begin{equation}
\label{eq: H}
{\cal H}[n_i]=\frac{1}{2} \sum_{ij} n_i V_{\rm SS} ({\bf r}_i-{\bf r}_j) n_j - \sum_i \mu n_i \ .
\end{equation}
Note that the sum is not over nearest neighbors but over all sites -- the range of the interaction is encoded in $V_{\rm SS}$. This model, with a square shoulder, was used to study electron liquids in weak magnetic fields~\cite{Shklovskii}. There it was a ``toy'' model of the true interactions, while here we study its consequences for generic $V_{\rm SS}$~\cite{remark1}.
To develop a mean-field theory, it behooves us to rewrite the partition function in terms of a continuous field $\phi$ through the Hubbard-Stratonovich transformation~\cite{remark2}
\begin{eqnarray}
{\cal Z} &=& \sum_{n_j}\!\int [{\rm d}\phi] \exp\!\!\Bigg\{\!-\frac{ \rho_0^2}{2 \beta}\int \frac{{\rm d}^d\!k}{(2\pi)^d} \phi({\bf k}) V^{-1}_{\rm SS}({\bf k})\phi(-{\bf k})\nonumber \\ &&\qquad+\sum_j n_j(i\phi_j+\beta \mu)\Bigg\} \ 
\end{eqnarray}
where we have mixed the continuum and discretum for notational ease and $d$ is the dimension of space. Here $\rho_{0}=\gamma_d \sigma^{-d}$ is the  density of the lattice where we have chosen the lattice constant to equal the hard sphere diameter $\sigma$ and  $\gamma_d$ is  a lattice and dimensionally dependent geometrical factor. Summing over $n_j=0,1$ results in an interacting Fermi-Dirac-like free energy.  The average site occupation is $\langle n_i \rangle = \beta^{-1} {\rm d}\ln{\cal Z}/{\rm d}\mu = \lambda e^{i \phi_i}/(1+\lambda e^{i  \phi_i})$ where $\lambda=e^{\beta \mu}$ is the fugacity.  This is related to the number density profile through $\rho_i=  \rho_{0} \langle n_i \rangle$.  Solving the mean-field equations, we find that $V^{-1}_{\rm SS}({\bf k})\phi({\bf k}) = i \beta \rho({\bf k})/\rho_0^2$ and the mean-field estimate for the fixed $\rho$ ({\sl not} fixed $\mu$) free energy, obtained
via Legendre transform, is the standard density functional form:
\begin{multline}
\beta F=\frac{\beta}{2\rho_0^2}\int \frac{{\rm d}^d\!k}{(2\pi)^d} \rho({\bf k})V_{\rm SS}({\bf k}) \rho(-{\bf k}) + \int {\rm d}^d\!r \big[ \rho\ln\left(\rho/\rho_{0} \right) \\ + \left(\rho_{0}-\rho\right)\ln\left(1-\rho/\rho_{0}\right)\big] \ . 
\end{multline}
Following Likos {\sl et al.}~\cite{Likos01}, we consider the stability of fluctuations around a uniform density. Working to quadratic order in fluctuations about a constant $\rho$ we find an instability when both $V_{\rm SS}^{-1}({\bf k})<0$ and $\rho_0\le \rho/\rho_{0} \left(1-\rho/\rho_{0} \right)\beta\vert V_{\rm SS}({\bf k})\vert$. At low volume fractions, when $\rho/\rho_{0} \rightarrow 0$ and the hard cores rarely overlap, we recover the result in Ref.~\cite{Likos01}. The presence of the hard core not only alters the stability criterion, but introduces a new length scale $\sigma$ which competes with the overlap of the shoulders at $\sigma_s$.

Though our lattice model provides a modicum of insight, it fails to distinguish ordered from disordered states in the clusters. Thus, we have analyzed the HCSS system off lattice in two dimensions by constructing a mean-field model which captures the salient features of the self-assembly. We consider the interaction of clusters of fixed shape in various periodic arrangements. The interaction between the clusters arises from the summed interactions of the spheres in each cluster, rotationally averaged so that we can ignore any spatial or orientational correlations between spheres in different clusters. It is then straightforward to calculate the average overlap energy for a given aggregate morphology. Moreover, for each morphology, we consider either fluid or crystalline order in the clusters; the entropy is estimated using an empirical equation of state for the hard-disk fluid~\cite{Santos95} or the cellular theory of the hard-disk crystal~\cite{Barker63}. The result is a closed, albeit cumbersome, analytic form as a function of temperature, density, and the two structural parameters, cluster size and lattice spacing (the latter are fixed via Lagrange multipliers). An example of the phase diagram of the aggregate phases, which are stable at low enough temperatures $\beta\epsilon\gtrsim1$, is shown in Fig.~\ref{fig:phasediagram}.

\begin{figure}[t]
\includegraphics{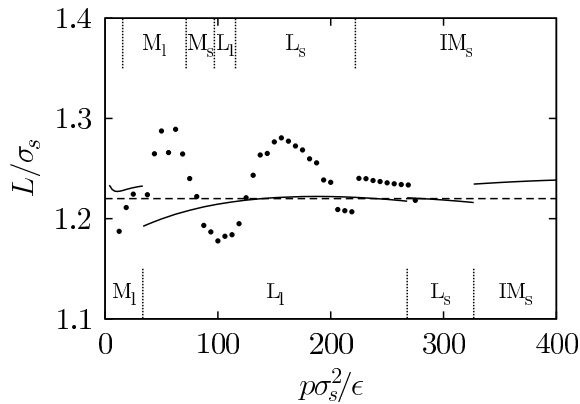}
\caption{Wavelength of principal mode vs. pressure for $\sigma_s/\sigma=5$ and $\beta\epsilon=2$.  The dashed line shows the prediction from the lattice, mean-field theory, $L/\sigma_s \simeq 1.22$, the solid line shows the results of the mean-field, off-lattice calculation with the boundaries of the respective phases labeled below; and the filled circles show the results from the simulations with the phases labeled above.  The phase morphologies are labeled as in Fig. \ref{fig:phasediagram}.  The lattice constant and the wavelength differ by a multiplicative factor of $\sqrt{3}/2$ in the hexagonal phases, as usual.}
\label{latticeconstant}
\end{figure}

In units of $\sigma_s^{-1}$, the wavenumber at which ordering occurs only depends weakly on the dimensionless pressure $p\sigma_s^2/\epsilon$ and is, interestingly, roughly independent of the particular phase, be it micellar, lamellar, or inverted micellar, as shown in Fig.~\ref{latticeconstant}. This behavior recapitulates the results for a generalized exponential interaction without hard cores~\cite{Mladek06}. Taken together with the observation that the
qualitative features of the phase diagram are independent of $\sigma_s/\sigma$, this confirms our notion that we can think of the clusters effectively as larger interacting particles with the internal structure only changing the effective cluster-cluster interactions. The relative unimportance of the hard-core interaction arises when the clusters are large and an additional sphere can be accommodated without requiring a large rearrangement of the internal structure. Moreover, we find that in the pressure and temperature ranges we consider, the micellar and lamellar phases can have both fluid and crystalline intra-aggregate order. In contrast, the only stable inverted micellar phase is crystalline.
Our analysis corroborates the results of Monte Carlo (MC) simulations of the HCSS system at fixed $N$, $p$,
and $T$~\cite{glaserprl}. At low temperatures, the MC simulations reveal a series of first-order transitions, with decreasing pressure, from crystalline inverse micelles, to crystalline lamellae, to liquid lamellae, to crystalline micelles, to liquid micelles, to isotropic liquid, in qualitative agreement with the phases and phase diagram of our mean-field treatment. Transitions between distinct structural variants of each type of phase (e.g., between different crystalline lamellar phases) are also observed. A more detailed discussion of the simulation results appears elsewhere~\cite{glaserprl}.

To understand the unwavering magnitude of the wavevector, we first consider the lamellar phase in the limit where the boundaries between the occupied and unoccupied regions are sharp, in the spirit of the strong segregation limit of diblock copolymers~\cite{GrasonReview}. Let $\ell$ be the width of the lamellae and $a$ be width of the gaps between them so that the centers of the lamellae are $L=\ell+a$ apart. If $\ell<\sigma_s$, the energy per sphere arising from other spheres in the same lamella is $E_{\rm intra}\approx\epsilon\rho_{\rm eff}\ell\sigma_s$ where $\rho_{\rm eff}=\rho(\ell+a)/\ell$ is the density of spheres in the occupied region. Thus $E_{\rm intra}\approx\epsilon\rho(\ell+a)\sigma_s$; a fixed lattice constant $L$ amounts to fixing the energy per particle arising from intra-aggregate interactions. Qualitatively, the inter-lamellar repulsion decreases both with $a$ and $\ell$:  as $a$ grows for fixed $\ell$, fewer particles in neighboring lamellae interact, and, as $\ell$ grows for fixed $a$, the fraction of the particles at the edge of each lamella which overlap more strongly with the neighboring lamella is reduced. To lowest order, our off-lattice mean-field analysis finds $E_{\rm inter}\approx0.75\epsilon\rho\sqrt{\sigma_s}(\sigma_s-a)^{5/2}(\ell+a)/\ell^2$, and the minimum of $E_{\rm inter}+E_{\rm intra}$ (subject to the condition that $\rho_{\rm eff}$ should not exceed the close-packed density $2/\sqrt{3}\sigma^2$) indeed lies on the line $\ell+a\approx1.2\sigma_s$. We also find that the equilibrium value of $\ell$ is as small as allowed by this condition, as if the lamellae were being compactified by an effective surface tension arising from the inter-lamellar repulsion. 

Returning to the lattice model, we evaluate the free energy of modulated phases; we need only consider Fourier modes which belong to the reciprocal lattice of the periodic structure (including the ${\bf k}=0$ mode). Fixing the zero mode of the density, we find that those wavevectors ${\bf k}$ for which $V_{\rm SS}({\bf k})<0$ are unstable, and that the strongest instability will be at the minimum ({\sl i.e., most negative}) value of $V_{\rm SS}$.
In the case of the soft shoulder in two dimensions, $V_{\rm SS}({\bf k}) =\epsilon\sigma_s J_1(k\sigma_s)/k$, which has a minimum at $k^*\sigma_s\approx2\pi/1.22$ in remarkably good agreement with the off-lattice mean-field results and the Monte Carlo simulations (Fig.~\ref{latticeconstant}). 

When is it appropriate to assume the strong-se\-gre\-ga\-tion regime and an infinitely narrow boundary
between occupied and unoccupied sites? To assess the accuracy of this approximation, we return to the mean-field treatment of the lattice model at {\it fixed chemical potential}. In the low temperature regime, we expect the boundaries to be infinitely narrow. In this limit, we can use a Sommerfeld-like expansion to compute $\rho({\bf r})$ for a stripe configuration to lowest order in $\Delta$, the width of aggregate boundary. As is the case for the square shoulder, we assume that $|V_{{\rm SS}} ({\bf k})|$ decays sufficiently rapidly with $|{\bf k}|$ and that the \textit{potential}, $\phi({\bf r})$, is well-approximated by $\phi({\bf r})\simeq \phi_0+\phi^*\cos(k^* x)$.
These coefficients are related by the mean-field equation to appropriate Fourier modes of the density.  At sufficiently low temperature the density is a step function, and the relevant Fourier modes are $\rho(k=0)/V_{tot}=\rho_{0} k^*\ell/2 \pi$ and $ \rho(k^*)/V_{tot} =\rho_{0} \sin(k^*\ell/2)/ \pi$, where $V_{tot}$ is the total volume (note that $\phi_0$ and $\phi^*$ are {\sl not} Fourier modes).  Simultaneously, the nonlinear relation between $\rho$ and $\phi$ gives an additional constraint on the position of the interface, $x=\ell/2$: $-\beta \mu=i\phi_0+i\phi^*\cos(k^*\ell/2)$.  Combining this with the mean-field equations,  we eliminate $\phi_0$ and $\phi^*$ to find a transcendental equation for $\ell$:
\begin{equation}
 \frac{V^*}{\pi} \sin k^* \ell = V_0 \frac{ k^* \ell}{ 2 \pi} - \mu \ ,
\end{equation}
where $V_0 \equiv  V_{{\rm SS}}(0)/\rho_0$ and $V^*\equiv  \vert V_{{\rm SS}} (k^*)\vert/\rho_0$. When we have a soft shoulder, $V_0 \gg V^*$ and $k^*\ell/(2\pi) \simeq \mu/V_0$. Temperature does not strongly affect the width of the aggregates, but it does alter the sharpness of inner domain boundary. The width of the interface is measured from the slope of the density at the interface, and is given by $\Delta^{-1} = \vert\rho^{-1} ({\rm d} \rho/{\rm d}x)\vert$ at $x=\ell/2$.  To lowest order, we find $k^* \Delta/(2 \pi) \simeq k_B T/[2V^* \sin^2(\pi \rho/\rho_{\rm eff})]$.
Therefore, we can see that the aggregates ``melt" from the boundaries as the temperature is raised due to the fact that the depth of the self-consistent potential is proportional to $V^*$.
We deduce a rough estimate of the stripe melting temperature, $T_m$, from the condition that $k^* \Delta/2 \pi \lesssim 1$ for ordered structures, or $k_B T_m \sim V^* \sin^2 (\pi \rho/\rho_{\rm eff})$.  From this we see that the most stable structures occur at {\it half-filling}, when $\rho_{\rm eff}=2\rho$, consistent with our instability criteria.
These calculations can be repeated for the other morphologies by utilizing more modes with $|\textbf{k}| = k^*$.

Our lattice model can be further embellished by including a short-range attraction in addition to a long-range repulsion~\cite{Tkachenko}. Though the details would be different, it is straightforward to see that our conclusion of an ordering instability would remain and that our qualitative arguments would still hold. Clearly, this aggregation phenomenon is encoded in the Fourier transform of the soft part of the pair potential, and we expect that self-assembly within a broad class of repulsive systems can be described by this model.
This analysis also suggests that the morphological similarities between repulsive self-assembly and clustering due to competing short-range attraction and long-range repulsions~\cite{seul95} are more than superficial: they reflect the common structure of the Fourier transform of the interaction potential.

In addition, we note that the lattice model, by virtue of its likeness to the Ising model, has a certain symmetry under toggling occupied and unoccupied sites (at half filling). This gives us a way to understand the existence of the inverse phases, and suggests that this symmetry is approximately preserved even in the off-lattice model. Finally, we note that the observed phase sequence is analogous to the ordered phases of diblock copolymers~\cite{Bates90} and amphiphiles in water~\cite{Israelachvili76}, despite the symmetry of the interaction. Near the critical point of the molten copolymer mean-field theory, quite generic Landau-theory considerations lead to the diversity of modulated structures in equilibrium \cite{leibler}.  We expect the structure of the mean-field theory for the present colloidal model to be very similar to the copolymer theory, at least near the mean-field critical point.
This suggests that, in three dimensions, HCSS systems can be expected to form spherical, cylindrical, and lamellar aggregates (and their inverses); indeed, such phases have been observed in MC simulations of the three-dimensional HCSS system~\cite{glaserprl}. Diblock melts also exhibit a bicontinuous gyroid phase, and it would be intriguing to identify an analogue for repulsive, self-assembling particles.

PZ was supported by Slovenian Research Agency through Grant P1-0055. CDS and RDK were supported through NSF Grant DMR05-47230, the Donors of the ACS Petroleum Research Fund, and a gift from L. J. Bernstein. GMG was supported through NSF Grant DMR04-04507. MAG was supported through NSF MRSEC Grant DMR02-13918. The authors thank the Aspen Center for Physics where this work was done.


\begin{thebibliography}{30}
\bibitem{Rabe} A. R. Kortan, N. Kopylov, R. M. Fleming, O. Zhou, F. A. Thiel, R. C. Haddon, and K. M. Rabe, \prb {\bf 47}, 13070 (1993).
\bibitem{AlexanderMcTague} S. Alexander and J. McTague, \prl {\bf 41}, 702 (1978).
\bibitem{Shklovskii} A. A. Koulakov, M. M. Fogler, and B. I. Shklovskii, \prl {\bf 76}, 449 (1996); M. M. Fogler, A. A. Koulakov, and B. I. Shklovskii, \prb {\bf 54}, 1853 (1996).
\bibitem{ZiherlKamienReview} P. Ziherl and R. D. Kamien, \prl {\bf 85}, 3528 (2000); J. Phys. Chem. B {\bf 105}, 10147 (2001).	
\bibitem{GrasonReview} G. M. Grason, Phys. Rep. {\bf 433} (2006) 1.
\bibitem{Torquato} M. Rechtsman, F. Stillinger, and S. Torquato, \pre {\bf 73}, 011406 (2006).
\bibitem{Young77} D. A. Young and B. J. Alder, \prl {\bf 38}, 1213 (1977).
\bibitem{Malescio03} G. Malescio and G. Pellicane, Nature Materials {\bf 2}, 97 (2003); \pre {\bf 70}, 021202 (2004).
\bibitem{Camp03} P. J. Camp, \pre {\bf 68}, 061506 (2003).
\bibitem{Norizoe05} Y. Norizoe and T. Kawakatsu, Europhys. Lett. {\bf 72}, 583 (2005). 
\bibitem{Likos01} C. N. Likos, A. Lang, M. Watzlawek, and H. L\"owen, \pre {\bf 63}, 031206 (2001).
\bibitem{Klein94} W. Klein, H. Gould, R. A. Ramos, T. Clejan, and A. I. Mel'cuk, Physica A {\bf 205}, 738 (1994).
\bibitem{Mladek06} B. M. Mladek, D. Gottwald, G. Kahl, M. Neumann, and C. N. Likos, \prl {\bf 96},
045701 (2006).
\bibitem{glaserprl} J. Santos, S. A. Kadlec, Z. V. Smith, J. Hausinger, P. D. Beale, N. A. Clark, and M. A. Glaser, submitted to \prl (2006).
\bibitem{remark1} Perhaps it was put best by Thoreau (in {\sl Walden}) , ``Our inventions are wont to be pretty toys, which distract our attention from serious things. They are but improved means to an unimproved end.''
\bibitem{remark2} Because it oscillates in sign, $V_{\rm SS}({\bf k})$ has a zero and is not invertible. We can, however, decompose $V_{\rm SS} = V_+ - V_-$ into nonvanishing potentials. We then introduce {\sl two} dummy scalar fields $\phi_\pm$ to complete the Hubbard-Stratonovich transformation. This does not alter the mean-field equations.
\bibitem{Santos95} A. Santos, M. L\'opez de Haro, and S. Bravo Yuste, \jcp {\bf 103}, 4622 (1995).
\bibitem{Barker63} J. A. Barker, {\sl Lattice Theory of the Liquid State} (Pergamon Press, 
Oxford, 1963).
\bibitem{Tkachenko} A. V. Tkachenko, \prl {\bf 89}, 148303 (2002).
\bibitem{seul95} See, for example, M. Seul and D. Andelman, Science {\bf 267}, 476 (1995) and references therein.
\bibitem{Bates90} F. S. Bates and G. H. Fredrickson, Ann. Rev. Phys. Chem. {\bf 41}, 525 (1990).
\bibitem{Israelachvili76} J. N. Israelachvili, D. J. Mitchell, and B. W. Ninham, J. Chem. Soc. 
Faraday Trans. 2, {\bf 72} 1525 (1976).
\bibitem{leibler}
L. Leibler, Macromolecules {\bf 13}, 1602 (1980).
\end{thebibliography}
\end{document}